\documentclass[aps,prb,twocolumn,showpacs]{revtex4-1}
\usepackage{graphicx}
\usepackage{natbib}
\usepackage[utf8]{inputenc}
\setcitestyle{super}

\begin{document}

\preprint{APS/123-QED}
\title{Combined effects of Sr substitution and pressure on the ground states in C\lowercase{a}F\lowercase{e}$_2$A\lowercase{s}$_2$}
\author{S. Knöner$^{1}$}
\author{E. Gati$^{1}$}
\author{S. Köhler$^{1}$}
\author{B. Wolf$^{1}$}
\author{U. Tutsch$^{1}$}
\author{S. Ran$^{2,\ddag}$}
\author{M. S. Torikachvili$^{3}$}
\author{S. L. Bud'ko$^{2}$}
\author{P. C. Canfield$^{2}$}
\author{M. Lang$^{1}$}
\address{$^{1}$Physikalisches Institut, J.W. Goethe-Universität Frankfurt(M), SPP1458, D-60438 Frankfurt(M), Germany}
\address{$^{2}$Ames Laboratory, US DOE, and Departement of Physics and Astronomy, Iowa State University, Ames, IA 50011, USA}
\address{$^{3}$Departement of Physics, San Diego State University, San Diego, California 92182, USA}
\address{$\ddag$ present address: Department of Physics, University of California, San Diego, California 92093, USA}
\date{\today}

\begin{abstract}
We present a detailed study of the combined effects of Sr substitution and hydrostatic pressure on the ground-state properties of CaFe$_2$As$_2$. Measurements of the electrical resistance and magnetic susceptibility, both at ambient and finite pressure $P$ $\leq$ 2\,GPa, were performed on Ca$_{1-x}$Sr$_x$Fe$_2$As$_2$ single crystals grown out of Sn flux. We find that upon Sr substitution the ranges of stability of both the structural-magnetic transition and the transition into the non-magnetic collapsed tetragonal phase are shifted to higher pressure levels with the latter moving at a higher rate. This suggests the possibility of separating the two phase lines, which intersect already at elevated temperatures for $x$ = 0 and low Sr concentration levels. For $x$ = 0.177 we find strong evidence that both phases remain separated down to lowest temperature and that a zero-resistance state emerges in this intermediate pressure window. This observation indicates that Sr-substitution combined with hydrostatic pressure provides another route for stabilizing superconductivity in CaFe$_2$As$_2$. Our results are consistent with the notion that (i) preserving the fluctuations associated with the structural-magnetic transition to low temperatures is vital for superconductivity to form in this material and that (ii) the non-magnetic collapsed tetragonal phase is detrimental for superconductivity.
\end{abstract}

\pacs{74.70.Xa, 74.25.fc, 74.62.Fj, 74.25.Dw}

\maketitle
\section{Introduction}

Among the parent compounds of Fe-based superconductors, CaFe$_2$As$_2$ manifests unique physical properties \cite{canfield2009}. The coupled magnetic/structural phase transition from the high-temperature tetragonal/paramagnetic (tet/pm) phase to the low-temperature orthorhombic/antiferromagnetic (o/afm) phase is strongly first order with a thermal hysteresis of several degrees Kelvin as seen in thermodynamic, transport and microscopic measurements\cite{ni2008b, goldman2008}. In addition, CaFe$_2$As$_2$ is the most pressure sensitive compound of the AeFe$_2$As$_2$ (122) or RFeAsO (1111) families\cite{torikachvili2008, kreyssig2008, goldman2009, yu2009, lee2009, prokes2010, park2008} where Ae = Ba, Sr, Ca and R = rare earth. The magnetic/structural phase transition at $T_{N}$/$T_{s}$ is initially suppressed by more than 100\,K per GPa, with the transition remaining strongly first order. As pressure increases above $\approx$ 0.35\,GPa, a nonmagnetic, collapsed tetragonal (cT) phase is stabilized. The cT phase boundary intersects and thereby terminates the low-pressure o/afm phase near 100\,K and 0.4\,GPa. The corresponding transition temperature $T_{cT}$ grows with increasing pressure and reaches 300\,K at a pressure of about 1.5\,GPa. In addition to its extraordinarily high pressure sensitivity, CaFe$_2$As$_2$ is highly sensitive to non-hydrostaticity the extent of which depends on the pressure medium used in the experiment\cite{canfield2009, yu2009, gati2012, taufour2014}. It was found that in pressure experiments, using oil as a pressure-transmitting medium, superconductivity shows up at low temperatures \cite{torikachvili2008}. In contrast, in experiments using $^4$He-gas as a pressure-transmitting medium, no superconductivity was observed\cite{yu2009}. The appearance of superconductivity has been assigned to anisotropic strains which are introduced into the material when the cT phase transition is crossed while the pressure medium is in its solid phase\cite{canfield2009,yu2009}. The reason for that is the large volume change of -5\% accompanied by a reduction of the ratio of the lattice parameters $c/a$ of -11\%\cite{kreyssig2008} upon cooling into the cT phase. Such an anisotropic strain leads to a multicrystallographic state with a small amount of strain-stabilized tetragonal phase which superconducts at low temperatures\cite{canfield2009, yu2009}. Since for a given pressure helium solidifies at much lower temperatures\cite{langer1961} as compared to oil, well below both the magnetic-structural and the cT transition in CaFe$_2$As$_2$, these non-hydrostatic effects are minimized in experiments using $^4$He-gas.


To further explore the propensity of CaFe$_2$As$_2$ to superconductivity, implied by these observations, Ran \textit{et al.} performed systematic studies of the effect of Co doping and annealing on single crystals grown out of FeAs flux \cite{ran2011, ran2012}. These studies suggested that the separation of the o/afm phase from the cT phase line is the key to revealing superconductivity in this material. It was demonstrated that, as a function of annealing temperature $T_{anneal}$, with increasing amount of Co the o/afm phase line is suppressed to lower temperatures whereas the cT phase stays roughly unchanged. By separating these two phase boundaries from each other superconductivity is stabilized. Later on it was shown that the annealing process mimics the effect of hydrostatic pressure\cite{gati2012}. Alternatively, one may expect that the separation of the two phase lines can also be realized in pressure experiments by moving the cT phase to higher pressures while keeping the o/afm phase line unchanged or at least moving it with a much lower rate. The fact that SrFe$_2$As$_2$ manifests a phase transition into the cT state at room temperature under 10\,GPa\cite{uhoya2011, kasinathan2011}, which is more than an order of magnitude higher than the pressure needed to stabilize the cT phase in CaFe$_2$As$_2$, is in support of this possibility. 
 Here we demonstrate that by partially substituting Ca by Sr in CaFe$_2$As$_2$ it is possible to separate the o/afm from the cT phase as a function of pressure, and by this, to stabilize a state of zero resistivity in the intermediate pressure range.

\section{Experimental details}

Single crystals of Ca$_{1-x}$Sr$_x$Fe$_2$As$_2$ were grown out of Sn flux using standard high-temperature solution growth techniques\cite{canfield2001, canfield1992} and were investigated via magnetic susceptibility and electrical resistance at ambient and finite pressures of $P$ $\leq$ 2\,GPa (resistance) and $P$ $\leq$ 0.5\,GPa (magnetic susceptibility). The crystals had plate-like shapes with several mm size in the $ab$-plane. Room temperature lattice parameters were determined by powder X-ray diffraction using a Rigaku Miniflex X-ray diffractometer. Single crystals were ground in liquid nitrogen so as to reduce deformations. The actual chemical composition was determined using wave-length dispersive X-ray spectroscopy (WDS) in a JEOL JXA-8200 electron microscope, by averaging twelve spots on the crystal surface.

For the measurements under pressure different pressure cells combined with different pressure media (pm) were used. Measurements of the magnetic susceptibility and electrical resistance under $^{4}$He-gas pressure were carried out for pressures $P$ $\leq$ 0.5\,GPa (magnetic susceptibility) and 1.1\,GPa (resistance). These measurements will be indicated by \textit{gas-pm} from now on and the crystals studied in this way are labeled by \#F. Resistance measurements using a clamp-type pressure cell were carried out with a mixture of either pentane/isopentane or oil/pentane under pressures up to 2\,GPa. We refer to these measurements as \textit{liquid-pm} and indicate the crystals studied in this way by \#A in the following. For the resistance measurements performed at \textit{liquid-pm} conditions, the pressure was generated in a Teflon cup which was filled with the pressure medium and which was inserted into a nonmagnetic, piston-cylinder-type CuBe pressure cell. In these experiments, the in-plane ac electrical resistance was measured by a standard four-probe configuration within a Quantum Design Magnetic Property Measurement System (MPMS) using a Linear Research LR700 resistance bridge ($f$ = 16\,Hz, $I$ = 1\,mA). The low-temperature pressure was determined by monitoring the shift of $T_c$ of elemental Pb whereas the room temperature value was determined by using a manganin wire as a manometer. Based on our experience with the cell used for these measurements, the uncertainties in the pressure values at intermediate temperatures (100\,K $\leq T \leq$ 260\,K) amount to approximately $\pm$ 50\,MPa.

For measurements of the in-plane electrical resistance and magnetic susceptibility under $^4$He-gas pressure, a CuBe pressure cell (Institute of High-Pressure Physics, Polish Academy of Sciences, Unipress Equipment Division) was used. The cell is connected by a CuBe capillary to a large volume He-gas compressor kept at room temperature to ensure $P$ $\simeq$ const. conditions during temperature sweeps. The susceptibility data, measured in a MPMS with $B$ $\parallel$ $ab$, have been corrected for the contribution of the sample holder, including the pressure cell, which was determined independently. The electrical resistance was measured in a four-terminal ac configuration by employing a Linear Research Bridge (LR700) ($f$ = 16 Hz, excitation voltage $U$ = 60\,$\mu$V) or with a dc technique using an HP3245A universal source ($I$ = 1\,mA) together with a Keithley 2182A nanovoltmeter. In electrical resistance measurements, an $n$-InSb\cite{kraak1984} single crystal was used for an $in$-$situ$ determination of the pressure. For susceptibility measurements, the shift of $T_c$ of elemental In was used for the determination of the pressure in the low-temperature regime \cite{jennings1958}. In addition, a manganin manometer on the compressor side was used at temperatures above the solidification of the pressure medium. The use of helium as a pressure-transmitting medium ensures truly hydrostatic pressure conditions as long as it is in the liquid phase, i.e., in a $T$ - $P$ range above the solidification line $T^{solid}_{He} (P)$. Even when cooling through $T^{solid}_{He} (P)$, which is accompanied by a pressure loss of about 30\%, deviations from hydrostatic conditions are relatively small. This is due to the low solidification temperature of helium, implying a small thermal expansion mismatch between sample and frozen pressure medium, and the small shear modulus of solid helium\cite{haziot2013}.
In order to reduce uncertainties related to potential changes of pressure as a function of temperature, in this report we indicate the pressure values which have been determined at a temperature close to the phase transition under discussion.

\section{Results}

\subsection{Characterization}

\begin{figure}[!t]
	\centering
		\includegraphics[width=0.98\columnwidth]{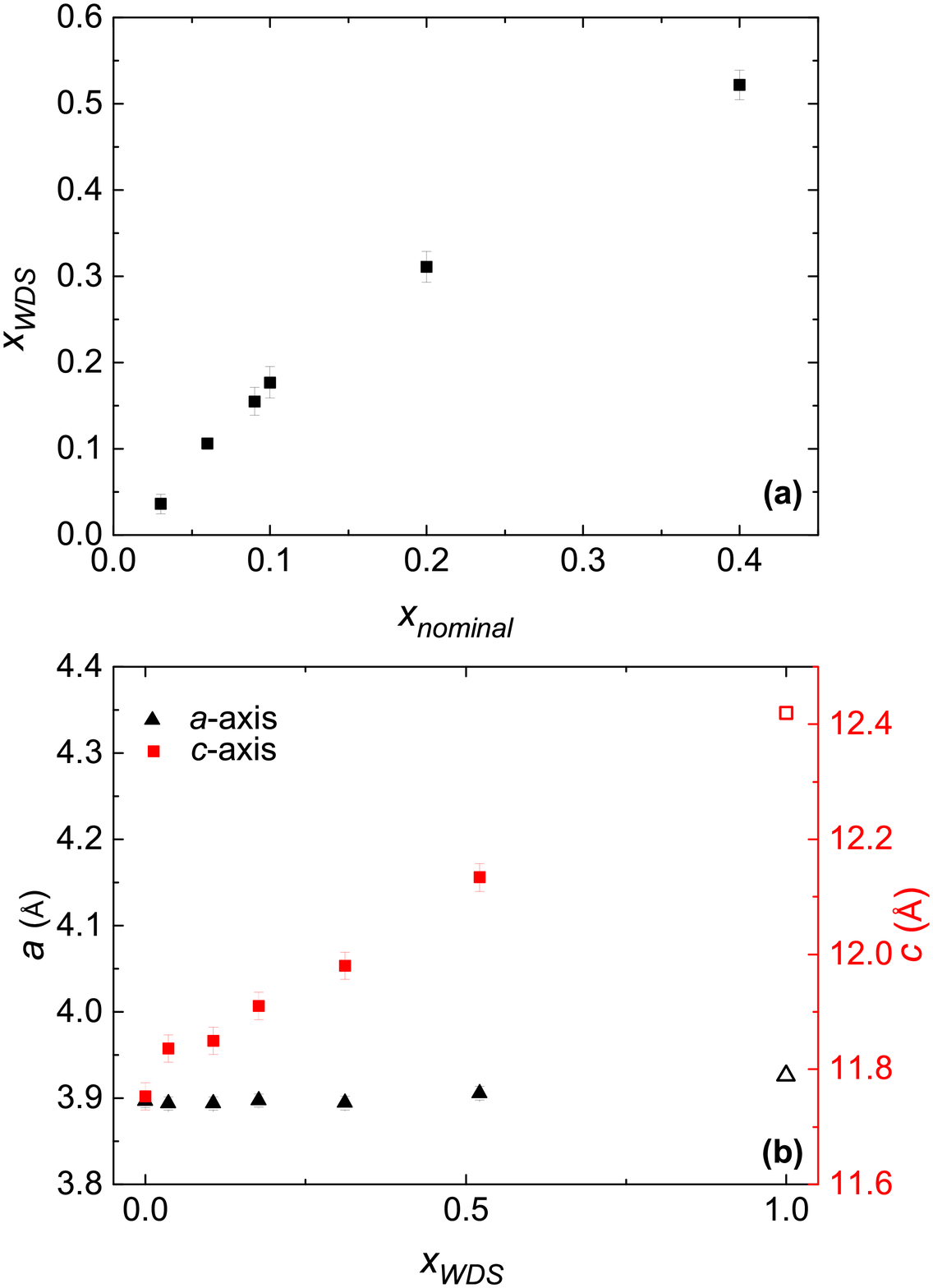}
		\caption{Results of the chemical analysis and lattice parameters of Ca$_{1-x}$Sr$_x$Fe$_2$As$_2$ as a function of $x$. (a) Measured Sr concentrations $x_{WDS}$ \textit{vs.} nominal concentration. (b) Lattice parameters $a$ (left axis, full black triangles) and $c$ (right axis, full red squares) as a function of $x_{WDS}$. Data for $x$ = 1 (open symbols) are taken from Yan \textit{et al.}\cite{yan2008}.}
	\label{fig:1}
\end{figure}

In Fig.\,\ref{fig:1}(a) we present the results of the WDS experiments. The figure demonstrates that, for $x_{nominal}$ $<$ 0.1, the data points of nominal versus actual concentration can be fit fairly well by a straight line with a slope of 1.73\,$\pm$0.07. With further increasing $x_{nominal}$, the actual Sr concentration progressively deviates from the straight line. The error bars are taken as twice the standard deviation determined from the measurements, and never exceed 0.02. This demonstrates the homogeneity of the substitution in the samples investigated. In the following, the averaged experimentally determined $x$ values, $x_{WDS}$, will be used to identify the compounds. In Fig.\,\ref{fig:1}(b) the $a$ and $c$-lattice parameters are shown as a function of Sr concentration $x_{WDS}$. The figure, which also includes literature data for pure SrFe$_2$As$_2$ ($x$ = 1)\cite{yan2008}, taken on the same X-ray machine, shows that the $a$-lattice parameter increases only very little with increasing Sr concentration, whereas the $c$-lattice parameter grows significantly in a monotonous way. These results are similar to those of Ref. \cite{saha2011} and demonstrate that the in-plane lattice parameters are governed by the FeAs layers, which are only weakly affected by Sr substitution, whereas the blocking layers grow in thickness by Sr substitution due to its larger ionic radius. Given that the cT phase originates from the formation of interlayer As-As bonds, as first pointed out by Hoffmann \textit{et al.} \cite{Hoffmann1985}, a larger $c$-lattice parameter will potentially require higher pressure to drive the system into the cT phase. Therefore, for the purpose of moving the cT phase to higher pressures, Sr substitution appears promising.
	
\begin{figure}[!t]
	\centering
		\includegraphics[width=0.98\columnwidth]{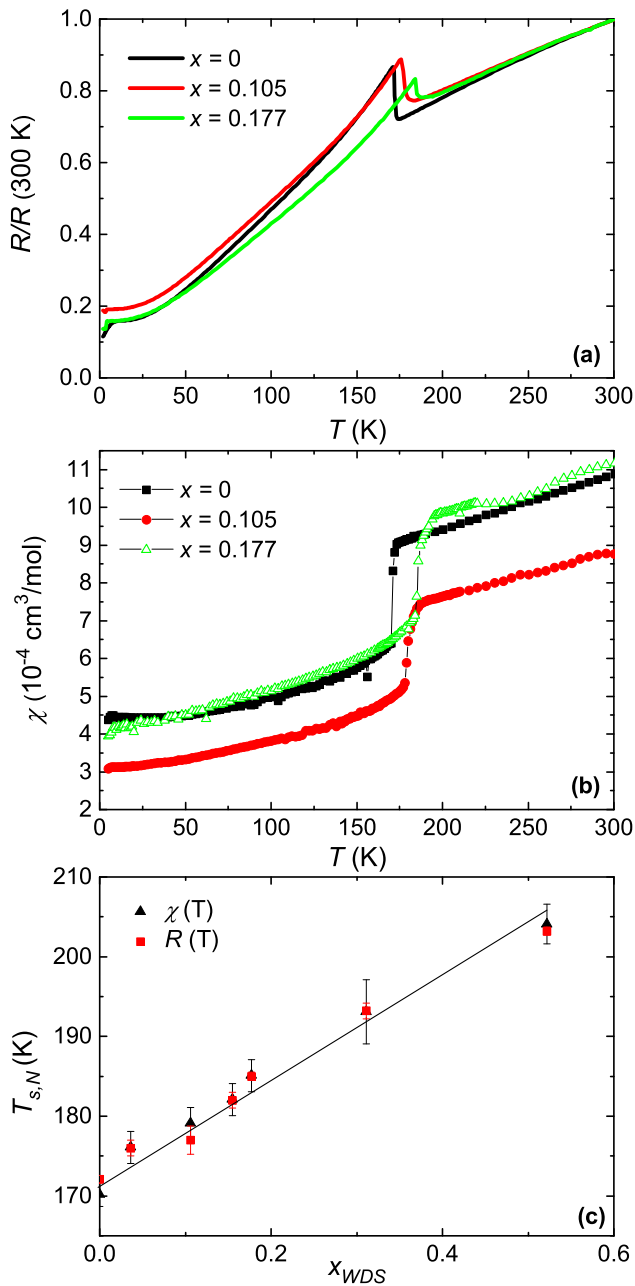}
		\caption {Temperature dependence of (a) the electrical resistance $R\,(T)$, normalized to its value at 300\,K, and (b) the magnetic susceptibility $\chi\,(T)$ for Ca$_{1-x}$Sr$_x$Fe$_2$As$_2$ single crystals with $x$ = 0, 0.105 and 0.177. (c) Transition temperature of the structural-magnetic phase transition at $T_{s,N}$ vs. $x_{WDS}$ for all concentrations prepared. Triangles correspond to magnetic susceptibility data ($\chi$), squares correspond to electrical resistance ($R$) data. The solid line is a guide to the eyes.}
	\label{fig:2}
\end{figure}

In Fig.\,\ref{fig:2}(a) we present the normalized temperature-dependent electrical resistance $R\,(T)$ and in Fig.\,\ref{fig:2}(b) we present the magnetic susceptibility $\chi\,(T)$ of samples with $x$ = 0, 0.105 and 0.177 at ambient pressure. The figure demonstrates that the form of the signatures in the resistance and magnetic susceptibility on cooling from the high-temperature tet/pm phase into the low-temperature o/afm phase, i.e., an abrupt increase of $R$ accompanied by a change in the slope, remain essentially unchanged with Sr substitution at this level. The observed jump-like anomalies are in accordance with the well-known signatures in these quantities revealed at the coupled first-order magnetic-structural transition in CaFe$_2$As$_2$ \cite{canfield2009}. The drop in the electrical resistance data at lowest temperatures in Fig.\,\ref{fig:2}(a) is a feature frequently observed in CaFe$_2$As$_2$ and SrFe$_2$As$_2$ samples, where it has been associated with surface-strain-induced filamentary superconductivity\cite{saha2009}. We assign the non-monotonic behavior of the absolute value of the magnetic susceptibility in Fig.\,\ref{fig:2}(b) to different amounts of flux contaminating the surface of the samples as well as differences in sample alignment. The data in Fig.\,\ref{fig:2}(a) and (b) clearly reveal an increase of the transition temperature $T_{s,N}$ with increasing Sr concentration. This is consistent with the fact that the magnetic-structural transition in pure SrFe$_2$As$_2$ occurs at an even higher temperature of 198\,K\cite{yan2008}. In Fig.\,\ref{fig:2}(c) we plot the $T_{s,N}$ data versus Sr concentration $x_{WDS}$. The figure also includes the transition temperatures derived in the same way for crystals with higher Sr concentration up to $x$ = 0.52, prepared in the course of this study. Figure\,\ref{fig:2}(c) demonstrates that up to $x \sim$ 0.5, to a good approximation, $T_{s,N}$ increases linearly with increasing $x_{WDS}$, even slightly exceeding the transition temperature for pure SrFe$_2$As$_2$. These data, unlike Ref. \cite{kirshenbaum2012}, imply that $T_{s,N}$ has a local maximum value for 0.52 $\leq x <$ 1.

\subsection{Pressure studies on crystals with $x$ = 0.105}

\begin{figure}[!t]
	\centering
	\includegraphics[width=0.96\columnwidth]{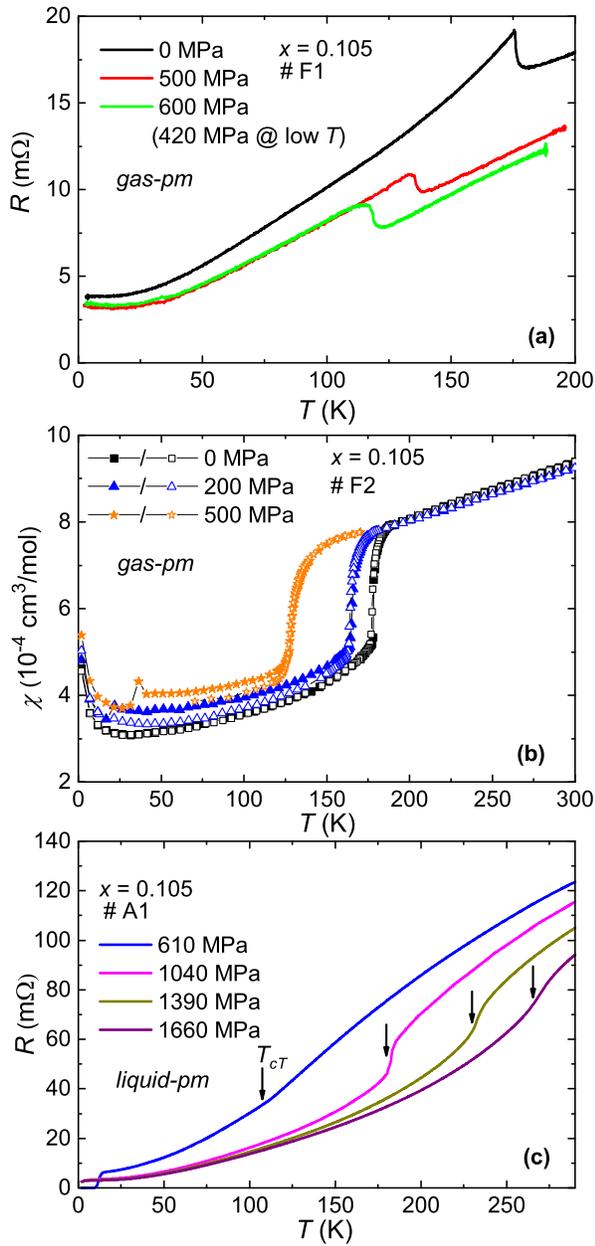}
		\caption {Single crystalline Ca$_{1-x}$Sr$_x$Fe$_2$As$_2$ for $x$ = 0.105 studied by (a) electrical resistance under \textit{gas-pm} conditions for selected pressures $P$ $\leq$ 600\,MPa on crystal \#F1 (data are taken on decreasing the temperature) and by (b) magnetic susceptibility under \textit{gas-pm} conditions for $P$ $\leq$ 500\,MPa on crystal \#F2. Closed symbols represent data taken with increasing temperature, open symbols data taken with decreasing temperature. The anomaly for $P$ = 500\,MPa at $T$ = 45\,K is assigned to solidification of helium. (c) Electrical resistance data for $x$ = 0.105 (crystal \#A1) under \textit{liquid-pm} conditions at higher pressure 610\,MPa $\leq$ $P$ $\leq$ 1660\,MPa taken on decreasing temperature. The arrows indicate positions of anomalies assigned to the transition into the cT phase.}
	\label{fig:3}
\end{figure}

The Figs.\,\ref{fig:3}(a) and \ref{fig:3}(b) show the temperature-dependent electrical resistance and magnetic susceptibility for crystals with $x$ = 0.105 measured under \textit{gas-pm} conditions at selected pressures $P$ $\leq$ 600\,MPa. The figure demonstrates that $T_{s,N}$ is reduced by the application of pressure. In addition, the shape of the magnetic-structural phase transition in both quantities remains essentially unchanged up to the highest pressures applied, indicating that the transition stays first order in the pressure range investigated. For the electrical resistance run labeled by $P$ = 600\,MPa, the pressure dropped to about 420\,MPa around 45\,K due to the solidification of helium. From the lack of any anomaly in $R$($T$) on crossing the solidification temperature and the fact that the data for $T\leq$ 100\,K, taken at $P$ = 500 and 600\,MPa, practically coincide, we infer that the resistivity in the o/afm phase is nearly pressure independent in this pressure regime. In contrast, the electrical resistance shows a pronounced decrease with increasing pressure in the tet/pm state, as observed previously in other Fe-based superconductors\cite{knoener2015, colombier2009, duncan2010, jeffries2012}. We note that no indications for superconductivity were found in either quantities under \textit{gas-pm} conditions.

\begin{figure}[!t]
	\centering
        \includegraphics[width=0.96\columnwidth]{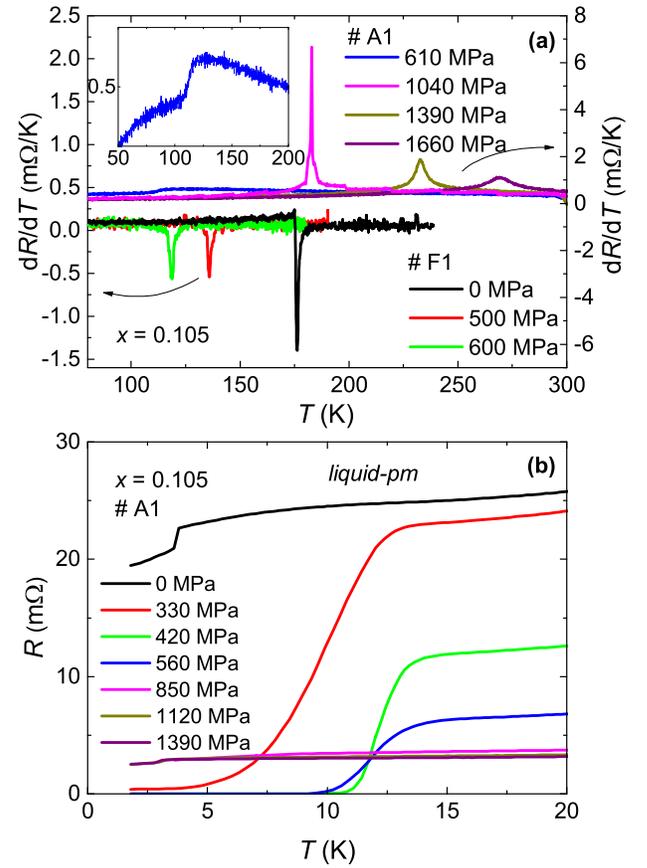}		
	\caption {(a) Anomalies in d$R$/d$T$ vs. $T$ for the $x$ = 0.105 data shown in Fig.\,\ref{fig:3}(a) corresponding to the structural-magnetic transition (left axis) and Fig.\,\ref{fig:3}(c) corresponding to the cT transition (right axis). The inset shows the anomaly around the transition at $P$ = 610\,MPa on enlarged scales. (b) Low-temperature part of the resistance data taken under \textit{liquid-pm} conditions. Note that the pressure value of 560\,MPa corresponds to the same measurement as $P$ = 610\,MPa in Fig.\,\ref{fig:3}(c).}
	\label{fig:4}
\end{figure}

Figure\,\ref{fig:3}(c) shows electrical resistance data for higher pressures 610\,MPa $\leq$ $P$ $\leq$ 1660\,MPa taken under \textit{liquid-pm} conditions. The arrows indicate the anomaly assigned to the transition from the high-temperature tet/pm into the low-temperature cT phase, see Fig.\,\ref{fig:4}(a) for the criterion used to determine the transition temperature. With increasing pressure, $T_{cT}$ shifts to higher temperatures, reaching $T_{cT} \approx$ 260\,K at 1660\,MPa. The figure also demonstrates that the signatures in $R$ at $T_{cT}$ change with pressure. In addition, for $P$ = 610\,MPa, the data reveal zero resistance at lowest temperatures. This will be further discussed in Fig.\,\ref{fig:4}(b). Unfortunately, measurements of the electrical resistance under \textit{gas-pm} conditions for $P\,>$ 600\,MPa failed to detect the phase transition into the cT phase due to the loss of electrical contacts, presumably caused by the large changes of the lattice parameters upon entering the cT phase. In contrast, for measurements performed under \textit{liquid-pm} conditions, where the pressure medium is already solid when the cT transition is crossed, no such problems occur. We tentatively assign this observation to a clamping effect the solid pressure medium exposes to the system, giving rise to reduced changes in the lattice parameters at the cT transition.

\begin{figure}[!t]
	\centering
		\includegraphics[width=0.98\columnwidth]{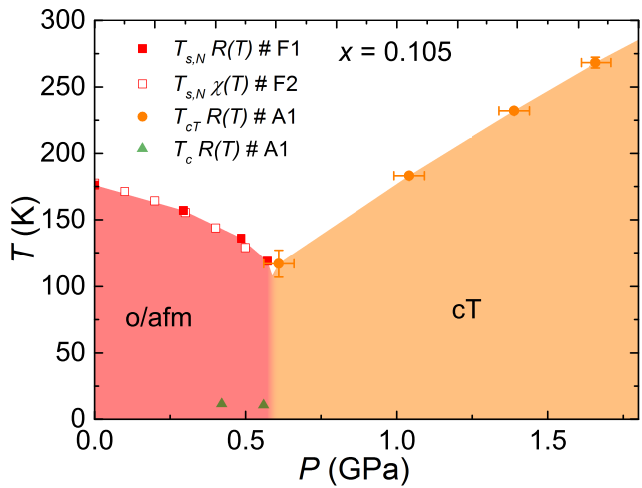}
		\caption {Temperature-pressure phase diagram of Ca$_{1-x}$Sr$_x$Fe$_2$As$_2$ for $x$ = 0.105. Closed (open) red squares correspond to the magnetic-structural transition temperatures, $T_{s,N}$, as seen in resistance (susceptibility) data taken under \textit{gas-pm} conditions, orange circles correspond to the transition to the cT phase seen in resistance measurements performed under \textit{liquid-pm} conditions. Green triangles indicate zero-resistance in measurements performed using \textit{liquid-pm}. The o/afm (cT) phase is marked by the red (orange) area.}
	\label{fig:5}
\end{figure}

For a closer inspection of the different phase transitions in the resistance shown in the Figs.\,\ref{fig:3}(a) and \ref{fig:3}(c), and for the purpose of defining criteria for the determination of the transition temperatures, we plot in Fig.\,\ref{fig:4}(a) the derivative of the electrical resistance $dR$/$dT$ vs. temperature. The magnetic-structural transition (left axis) is characterized by sharp minima in $dR$/$dT$ for all pressures studied. In contrast, the transition to the cT phase (right axis) manifests itself in a maximum in $dR$/$dT$, similar to the observation made in previous studies \cite{ran2011, ran2012}. We assign the position of the maximum in $dR$/$dT$ to the transition temperature $T_{cT}$. This includes also the small anomaly revealed for $P$ = 610\,MPa, which is shown in the inset to Fig.\,\ref{fig:4}(a) on enlarged scales. The assignment of this rather weak signature to a transition into the cT phase is corroborated by the position of the so-derived $T_{cT}$ value in the $T$-$P$ phase diagram in Fig.\,\ref{fig:5}, where we find a smooth continuation of the $T_{cT}$ values revealed at higher pressures.

Figure\,\ref{fig:4}(b) presents the low-temperature electrical resistance obtained under \textit{liquid-pm} conditions. The data for $P$ = 330\,MPa reveal a drop of the electrical resistance below $T$ $\approx$ 10\,K without reaching $R$ = 0 at lower temperatures. However, zero resistance is observed by increasing the pressure to $P$ = 420 and 560\,MPa. Note that the pressure value of 560\,MPa, determined at low temperatures $T$ $\leq$ 7 K, correspond to $P$ = 610\,MPa at elevated temperatures $T \approx$ 110\,K (see Fig.\,\ref{fig:3}(c)). Upon further increasing the pressure to 850, 1120 and finally 1390\,MPa, the zero resistance vanishes again. We stress, that measurements under \textit{gas-pm} conditions at $P$ = 420\,MPa (cf. Fig.\,\ref{fig:3}(a)) failed to reveal indications for superconductivity for $T \geq$ 2\,K.

Based on the resistance and susceptibility data in Figs.\,\ref{fig:3} and \ref{fig:4}, the phase diagram shown in Fig.\,\ref{fig:5} can be constructed. The transition temperatures of the magnetic-structural transition, $T_{s,N}$, extracted from resistance and susceptibility measurements under \textit{gas-pm} conditions match very well. We also find that the transition temperatures $T_{s,N}$ derived from experiments carried out under \textit{liquid-pm} conditions (not shown) agree to a good approximation with those of the \textit{gas-pm} experiments. Based on this observation, and in view of the lack of \textit{gas-pm} data for $T_{cT}$, we include in the phase diagram $T_{cT}$ values obtained under \textit{liquid-pm} conditions. Figure\,\ref{fig:5} demonstrates that the resulting $T_{s,N}$ and $T_{cT}$ lines meet at $\approx$ 600\,MPa and 110\,K, which is slightly different compared to the intersection point of $\approx$ 400\,MPa and 100\,K revealed for $x$ = 0. The phase diagram also includes the temperatures (green triangles) below which zero-resistance appeared in the measurements under \textit{liquid-pm} conditions. As will be discussed below, though, the zero-resistance state observed here is most likely associated with a minority phase caused by non-hydrostatic conditions.

\subsection{Pressure studies on crystals with $x$ = 0.177}

\begin{figure}[!t]
	\centering
		\includegraphics[width=0.98\columnwidth]{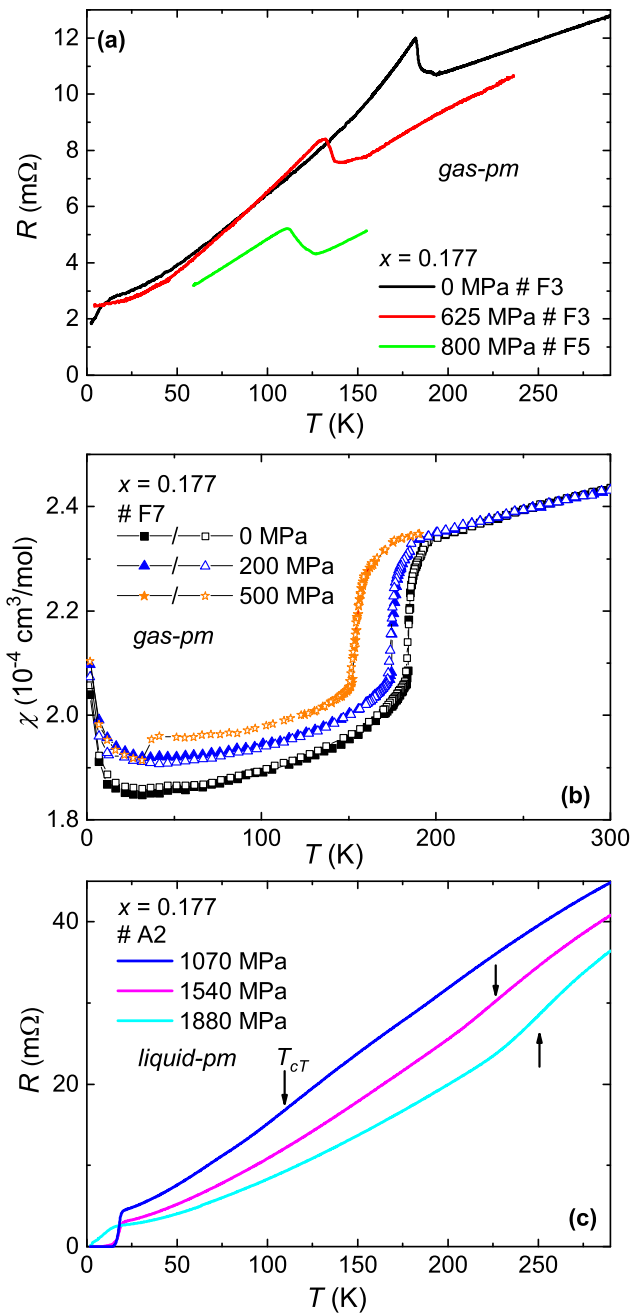}
		\caption {Single crystalline Ca$_{1-x}$Sr$_x$Fe$_2$As$_2$ for $x$ = 0.177 studied by (a) resistance measurements under \textit{gas-pm} conditions for selected pressures $P$ $\leq$ 800\,MPa on two different crystals upon decreasing temperature and by (b) magnetic susceptibility under \textit{gas-pm} conditions at $P$ $\leq$ 500\,MPa. Closed symbols represent data taken with increasing temperature, open symbols represent data taken with decreasing temperature. (c) Resistance data taken under \textit{liquid-pm} conditions at higher pressure 1070\,MPa $\leq$ $P$ $\leq$ 1880\,MPa, on crystal \#A2 upon decreasing temperature. The arrows indicate the position of anomalies assigned to the transition into the cT phase.}
	\label{fig:6}
\end{figure}

In the Figs.\,\ref{fig:6}(a) and \ref{fig:6}(b) we present the temperature-dependent resistance and magnetic susceptibility under \textit{gas-pm} conditions for crystals with $x$ = 0.177 at selected pressures $P$ $\leq$ 800\,MPa. Note that for the measurements at 800\,MPa (Fig.\,\ref{fig:6}(a)) crystal \#F5 was used whereas the experiments at 0 and 625\,MPa were carried out on crystal \#F3. To compensate for differences in the absolute resistances of these crystals, the data at 800\,MPa were multiplied by a factor to make the data fit into the scale used in this figure. Like for the $x$ = 0.105 crystal, the shape of the anomaly at the o/afm transition in both quantities does not change significantly with pressure. The data show that the magnetic-structural transition stays first order up to highest pressures investigated. Consistent with the observation made for the $x$ = 0.105 crystal, we find a decrease of $T_{s,N}$ with increasing pressure also for $x$ = 0.177.

\begin{figure}[!t]
	\centering
        \includegraphics[width=0.98\columnwidth]{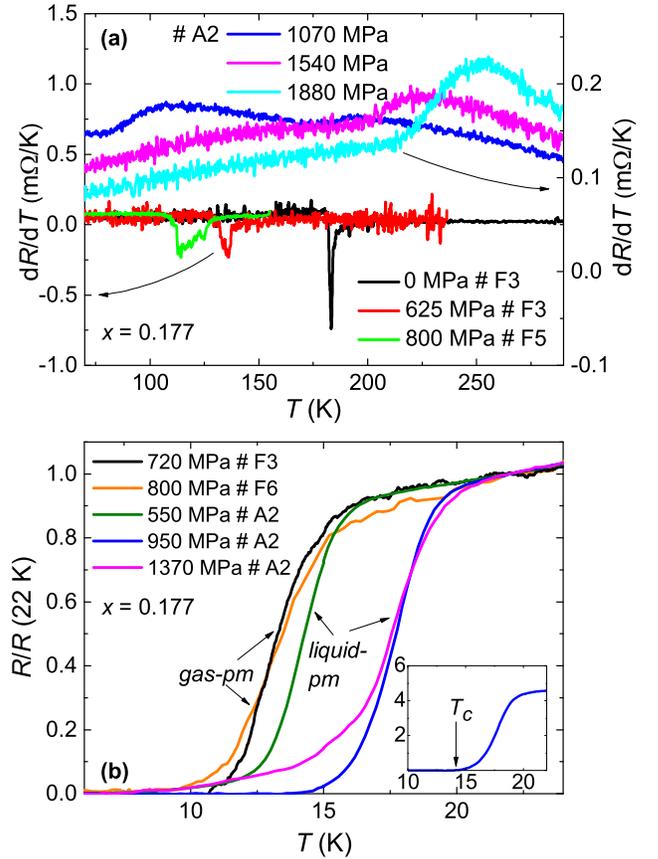}		
		\caption {a) Anomalies in d$R$/d$T$ vs. $T$ for the $x$ = 0.177 data shown in Fig.\,\ref{fig:6}(a) corresponding to the structural-magnetic transition (left axis) and \ref{fig:6}(c) the cT transition (right axis). b) Low-temperature part of the resistance, normalized to its value at 22\,K, under \textit{liquid-pm} conditions as well as \textit{gas-pm} conditions, measured on different crystals. Note that the pressure values of 950 and 1370\,MPa correspond to the same measurements as $P$ = 1070 and 1540\,MPa in Fig.\,\ref{fig:6}(c). The inset shows the ($R$ = 0) criterion used to determine the superconducting transition temperature.}
	\label{fig:7}
\end{figure}

Figure\,\ref{fig:6}(c) shows resistance data at higher pressures 1070\,MPa $\leq$ $P$ $\leq$ 1880\,MPa obtained under \textit{liquid-pm} conditions. In this pressure range we find similar anomalies as seen for the $x$ = 0.105 crystal which can be assigned to the transition to the cT phase. The arrows indicate the position of the maxima revealed in the derivative $dR$/$dT$, cf. Fig.\,\ref{fig:7}(a). With increasing Sr concentration the transition temperature $T_{cT}$ is shifted to higher temperatures. Furthermore, for $P$ = 1070 and 1540\,MPa, zero resistance was observed at low temperatures. This will be further considered in Fig.\,\ref{fig:7}(b).

In Fig.\,\ref{fig:7}(a) we present the derivative of the resistance $dR$/$dT$ of the data shown in the Figs.\,\ref{fig:6}(a) and \ref{fig:6}(c). We find that the anomaly corresponding to the magnetic-structural transition at $T_{s,N}$ (left axis) manifests itself in a sharp minimum in $dR$/$dT$, whereas the transition to the cT phase (right axis) is characterized by a broad maximum in $dR$/$dT$, consistent with the observations made for $x$ = 0.105. For the measurement at $P$ = 1070\,MPa the dominant feature is observed at $\approx$ 110\,K, which we assign to the cT transition. The data reveal yet another anomaly, of distinctly smaller size around 180\,K, the origin of which is unclear at present.

Figure\,\ref{fig:7}(b) presents the low-temperature resistance data for measurements performed under \textit{liquid-pm} (\#A2) as well as \textit{gas-pm} (\#F3, \#F6) conditions, labeled accordingly in the figure. The data at lowest ($P$ = 550\,MPa) and highest pressure (1370\,MPa) reveal rather long tails before reaching zero resistance. In contrast for intermediate pressure values $P$ = 720, 800 and 950\,MPa, the drop to $R$ = 0 is distinctly steeper, still having a width (10\% - 90\%) of $\approx$ 3\,K. Note that the pressure values of 950 and 1370\,MPa, determined at low temperatures $T$ $\leq$ 7 K, correspond to respectively $P$ = 1070 and 1540\,MPa at higher temperatures $T \geq$ 110\,K (see Fig.\,\ref{fig:6}(c)). We assign the transition temperature $T_c$ to the point below which zero resistance (within the experimental error) is observed, cf. the inset to Fig.\,\ref{fig:7}(b).

\begin{figure}[!t]
	\centering
		\includegraphics[width=0.98\columnwidth]{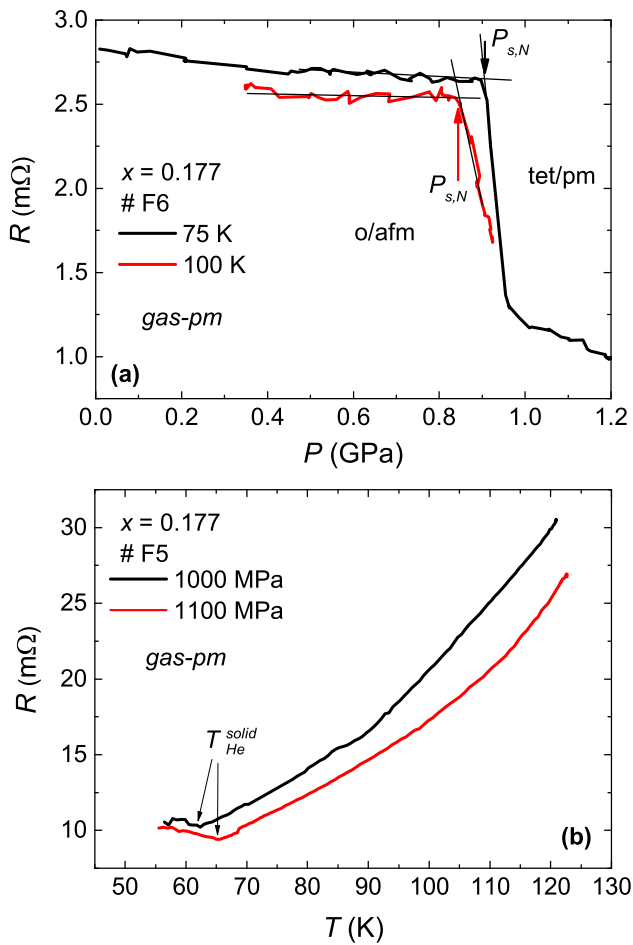}
		\caption {(a) Resistance data of Ca$_{1-x}$Sr$_x$Fe$_2$As$_2$ for $x$ = 0.177 taken at constant temperature $T$ = 75 and 100\,K with increasing pressure under \textit{gas-pm} conditions. The arrows indicate the onset pressure of the transition from the o/afm to the tet/pm phase. (b) $T$-dependent measurements at $P$ = 1000 and 1100\,MPa under \textit{gas-pm} conditions. The change of slope at $\approx$ 60\,K (65\,K) is due to the solidification of helium.}
	\label{fig:8}
\end{figure}

Figure\,\ref{fig:8}(a) shows resistance data taken at constant temperatures of $T$ = 100 and 75\,K upon increasing pressure under \textit{gas-pm} conditions. Note that for these temperatures and the whole pressure range shown there, $P \leq$ 1.2\,GPa, the pressure medium is in its liquid phase, eliminating any clamping effects and ensuring truly hydrostatic-pressure conditions. The data reveal a weak reduction of $R$ with increasing pressure up to about 0.8\,GPa. For the data taken at $T$ = 75\,K, this behavior is followed by a rapid drop in $R$ within a narrow pressure interval of $\Delta P \approx$ 50\,MPa with an onset pressure of (0.91 $\pm$ 0.03)\,GPa. For higher pressures $P >$ 0.95\,GPa up to the maximum accessible pressure of 1.2\,GPa, the resistance decreases moderately strongly with increasing pressure at a rate which is slightly higher than the one revealed for $P \leq$ 0.8\,GPa. A similar behavior is observed for the pressure sweep performed at $T$ = 100\,K, although with a somewhat reduced onset pressure of (0.845 $\pm$ 0.03)\,GPa. Note that the occurrence of a leak in the pressure system, causing a rapid loss of pressure, precluded measurements at higher pressures $P >$ 0.92\,GPa. We assign the rapid drop in the resistance to the first-order transition from the low-pressure o/afm phase to the high-pressure tet/pm phase and, in order use the same criterion for both runs, refer to the onset pressure as $P_{s,N}$. The discontinuous, slightly broadened change in $R$ to lower values on driving the system from the o/afm to the tet/pm phase is consistent with the jump-like resistance changes revealed in the temperature-dependent runs shown in Fig.\,\ref{fig:6}(a). The assignment of this drop in the resistance to the o/afm to the tet/pm transition is also corroborated by the fact that the electrical contacts stayed intact in these runs: given the frequently observed loss of contacts in our experiments under \textit{gas-pm} conditions on entry the cT phase, the intactness of the contacts is a strong indication that no phase boundary into the cT phase had been crossed in these runs. The small steps seen in the data with some hysteretic behavior are caused by the thermalization processes after a stepwise increase of the pressure. The different slopes of the data revealed at low and high pressure demonstrate that the resistance in the o/afm phase is only weakly pressure dependent, whereas it has a somewhat higher pressure dependence in the tet/pm phase. This is consistent with the observations made in the temperature-dependent measurements for $x$ = 0.105 (see Fig.\,\ref{fig:3}(a)) and $x$ = 0.177 (Fig.\,\ref{fig:6}(a)).

In Fig.\,\ref{fig:8}(b) we show the temperature-dependent resistance $R\,(T)$ measured at $P$ = 1000 and 1100\,MPa for temperatures 50\,K $\leq$ $T$ $\leq$ 120\,K, taken under \textit{gas-pm} conditions. The data lack any indications for a phase transition in the temperature range investigated. The change of slope at $\approx$ 60\,K ($P$ = 1000\,MPa) and 65\,K (1100\,MPa) is due to the pressure loss caused by the solidification of helium occurring at the temperatures indicated by arrows in the figure. The increase of the resistance upon further cooling below the solidification temperature is consistent with the notion that the sample in this temperature and pressure range is in the tet/pm state, given the negative pressure coefficient of the resistance inferred for this state from Fig.\,\ref{fig:8}(a). At temperatures below 55\,K, no data are available due to a loss of electrical contacts, presumably caused by the solidification of the pressure medium.

\begin{figure}[!t]
	\centering
		\includegraphics[width=0.98\columnwidth]{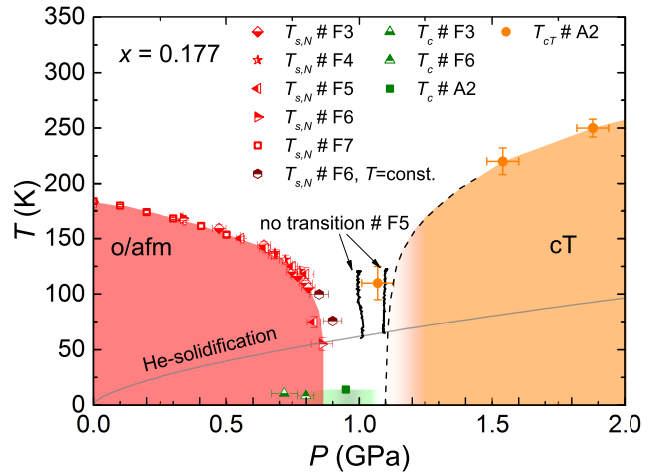}
		\caption {Temperature-pressure phase diagram of Ca$_{1-x}$Sr$_x$Fe$_2$As$_2$ for $x$ = 0.177. Red symbols mark the phase transition line from the tet/pm to the o/afm phase as determined under \textit{gas-pm} conditions: light red symbols correspond to $T_{s,N}$, determined from temperature-dependent measurements, whereas the dark red hexagons correspond to $P_{s,N}$ obtained from pressure-dependent measurements at constant temperature. Full orange spheres mark the transition to the cT phase as obtained under \textit{liquid-pm} conditions. Green symbols indicate the transition to a state of zero resistance. Solid black lines at $P$ = 1000 and 1100\,MPa correspond to temperature-pressure traces underlying the experiments shown in Fig.\,\ref{fig:8}(b), where no indications for a phase transition were found. The solid gray line shows the solidification line of helium.}
	\label{fig:9}
\end{figure}
The position of the various phase transitions revealed from the resistance and susceptibility measurements on $x$ = 0.177 are summarized in the phase diagram in Fig.\,\ref{fig:9}. For the magnetic-structural transition we find a monotonous suppression of the transition temperature $T_{s,N}$ upon increasing pressure with only little scatter in $T_{s,N}$($P$) derived from the various experimental probes at $P$ = const. on different crystals. Small deviations in the position of the phase transition are visible when comparing $T_{s,N}$($P$) with $P_{s,N}$($T$) results (marking the onset pressure) from the $T$ = const. experiments shown in Fig.\,\ref{fig:8}(a). We assign these discrepancies to the strongly first-order character of the structural-magnetic transition. Since $T_{s,N}$($P$) at zero pressure for $x$ = 0.177 is slightly enhanced, compared to $x$ = 0.105, somewhat higher pressures are needed here to suppress the o/afm phase.
The phase diagram also includes the transition temperature into the cT phase, $T_{cT}$($P$), derived from resistance measurements under \textit{liquid-pm} conditions. Due to the peculiarities of this pressure technique, less data points with larger error bars are available here. The data reveal the same trend as seen for the $x$ = 0.105 crystal, namely an increase of $T_{cT}$($P$) with increasing pressure. Most importantly, however, we find that the cT phase is shifted further away from the o/afm phase on the pressure axis as compared to the $x$ = 0.105 system. This implies the possibility that at intermediate pressures both phases remain separated down to lowest temperatures.

For a closer inspection of this intermediate pressure range, we include in Fig.\,\ref{fig:9} the temperature-pressure traces (solid black lines) underlying the experiments shown in Fig.\,\ref{fig:8}(b). As discussed above, the results of these resistance measurements, performed under \textit{gas-pm} conditions at $P \simeq$ 1000 and 1100\,MPa, suggest that in the temperature and pressure range covered by these experiments (i) the crystal is in the tet/pm phase and (ii) no phase transition line has been crossed. These observations would imply the $T_{cT}$($P$) phase boundary to proceed in a way as indicated by the broken line in Fig.\,\ref{fig:9}, still being compatible with the error bar of the $T_{cT}$ point centred at 1.07\,GPa. The rapid drop of the broken line at a pressure $P <$ 1.1\,GPa, implying a considerably wide pressure window where the system is in the tet/pm phase down to lowest temperatures,   is consistent with the results in Fig.\,\ref{fig:8}(a) from which we conjectured that at $T$ = 75\,K and $P \leq$ 1.2\,GPa, no phase boundary to the cT was crossed.

As for superconductivity, we include in the phase diagram only those data points where zero resistance combined with a relatively sharp transition (transition width of about 3\,K) was observed. This is the case for the crystals \#F3, \#F6 and \#A2 measured under \textit{gas-pm} (\#F3, \#F6) and \textit{liquid-pm} (\#A2) conditions, respectively. We refrain from including the data points taken at $P$ = 550 and 1370\,MPa (Fig.\,\ref{fig:7}(b)), where the transition is characterized by a rather long tail towards low temperatures. We consider this as an indication for inhomogeneous superconductivity, likely caused by non-hydrostatic pressure components. Such non-hydrostatic-pressure effects cannot be fully excluded even for measurements under \textit{gas-pm} conditions, since for the pressure range of interest, helium is already in its solid phase.

\section{Discussion}

\begin{figure}[!t]
	\centering
		\includegraphics[width=0.98\columnwidth]{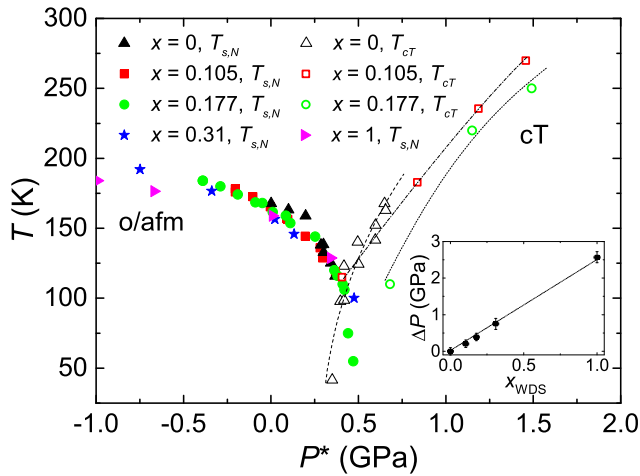}
		\caption {$T$-$P^*$ phase diagram for Ca$_{1-x}$Sr$_x$Fe$_2$As$_2$ for various $x$. Closed (open) symbols represent the transition to the o/afm (cT) phase. Black triangles correspond to data obtained on crystals with $x$ = 0, red squares to $x$ = 0.105, green circles to $x$ = 0.177, blue stars to $x$ = 0.31 and purple triangles to $x$ = 1. The inset shows the shift $\Delta P$ as function of concentration $x$ as described in the text. The dashed, respectively dottet lines are guide to the eyes. Data for $x$ = 0 and 1 are taken from Yu \textit{et al.}\cite{yu2009} and Colombier \textit{et al.}\cite{colombier2009}.}
	\label{fig:10}
\end{figure}

In discussing the combined effects of Sr substitution and pressure on the ground states in Ca$_{1-x}$Sr$_x$Fe$_2$As$_{2}$, we will proceed as follows. In a first step, we analyze to what extent the Sr substitution acts as chemical pressure. To this end, we shift the $T$-$P$ phase diagrams for $x$ = 0.105 and 0.177 by a pressure $-\Delta P$($x$) on the pressure axis such that the $T_{s,N}$($P-\Delta P$($x$)) lines match that of the $x$ = 0 system. The corresponding $T_{s,N}$($P$) data taken under \textit{gas-pm} conditions for $x$ = 0 are taken from Yu \textit{et al.} \cite{yu2009}. We also include in this analysis $T_{s,N}$($P$) data for $x$ = 0.31 derived from resistance measurements (not shown) and literature data for $x$ = 1 \cite{colombier2009}, both obtained under \textit{liquid-pm} conditions. The anomaly at $T_{s,N}$ revealed for the $x$ = 0.31 crystal is very similar to that observed under \textit{gas-pm} conditions for $x$ = 0.105 (Fig.\,\ref{fig:3}) and 0.177 (Fig.\,\ref{fig:6}(a)), albeit some broadening is visible with increasing pressure. The resulting $T$-$P^*$ phase diagram, with $P^*$ = $P$ - $\Delta P$, is presented in Fig.\,\ref{fig:10}. In the inset of this figure we show the required shift $\Delta P$($x$) for the various substitution levels $x$. The figure demonstrates that, to a good approximation, $\Delta P$ grows linearly with $x$ up to $x$ = 1. This demonstrates that regarding the structural magnetic-transition at $T_{s,N}$, Sr-substitution acts as chemical pressure. At the same time, the figure also demonstrates that this is not true for the cT transition line which becomes progressively shifted to higher pressures $P^*$ on increasing $x$ to 0.105 and 0.177. For $x$ = 0.31 no indications for a transition into the cT phase could be observed in our measurements (not shown) up to 1.8\,GPa. Taken together, while the o/afm phase shifts linearly to higher pressures with increasing Sr-substitution, the cT phase moves at a higher rate, likely in a non-linear fashion. We recall that the cT phase for $x$ = 1 shows up at $P$ = 10\,GPa, corresponding to $P^*$ about 7.5\,GPa, which is way off the scale used in Fig.\,\ref{fig:10}. These observations verify the idea of a pressure-induced separation of the o/afm and the cT phase for Sr-substituted CaFe$_2$As$_2$.

For $x$ = 0.105 and $x$ = 0.177, our studies reveal indications for pressure-induced superconductivity. The phase diagram obtained for $x$ = 0.105 in Fig.\,\ref{fig:5} demonstrates that the $T_{cT}$ phase line truncates the $T_{s,N}$ line at elevated temperatures. This scenario is similar to the situation encountered in pure CaFe$_2$As$_2$ where both phases intersect at approximately 100\,K and where the appearance of superconductivity has been assigned to non-hydrostatic pressure components \cite{torikachvili2008, yu2009}. We think that a similar scenario is existent here as measurements at 420\,MPa under \textit{gas-pm} conditions failed to reproduce the zero-resistance results obtained under \textit{liquid-pm} conditions. For that reason we conclude, that the observed indications for superconductivity for $x$ = 0.105 do not represent real bulk superconductivity, but are likely due to strain effects resulting from non-hydrostatic pressure conditions.

On the other hand, for the $x$ = 0.177 system, the $T$-$P$ phase diagram in Fig.\,\ref{fig:9} suggests that the situation here is distinctly different. For this concentration level there is evidence for a finite separation of the o/afm and the cT phases down to lowest temperatures and the emergence of superconductivity in the resulting gap between these two phases. Due to the lack of a volume-sensitive probe, however, we cannot make a clear statement regarding the bulk character of the observed superconducting state at present. But at least the $P$ = 950\,MPa region seems likely to support bulk superconductivity.

We stress that for both concentrations $x$ = 0.105 and 0.177, no indications for superconductivity are found on entering the cT phase at higher pressures. This is consistent with the observations made for Co-substituted CaFe$_2$As$_2$ studied under \textit{gas-pm} conditions \cite{gati2012} where superconductivity was found to disappear abruptly at the onset of the cT phase.

The above results demonstrate that isoelectronic Sr substitution in CaFe$_2$As$_2$ modifies the way pressure acts on the structural-magnetic transition at $T_{s,N}$ and the transition into the collapsed tetragonal phase at $T_{cT}$. Our investigations on $x$ = 0.177 reveal a finite pressure window where the o/afm transition has been fully suppressed and no cT transition interferes, allowing superconductivity to emerge. These observations are consistent with the notion that (i) preserving the correlations/fluctuations associated with the o/afm transition to low temperatures is vital for superconductivity to form in this material and that (ii) the cT phase is detrimental for superconductivity.\\

\section{Summary}

We have presented a detailed study on the combined effects of isovalent Sr-substitution and hydrostatic pressure on the ground states of CaFe$_2$As$_2$. Measurements of the electrical resistance and the magnetic susceptibility have been performed on various Sn-flux grown Ca$_{1-x}$Sr$_x$Fe$_2$As$_2$ single crystals both at ambient and finite pressure by using different pressure media. Measurements at ambient pressure reveal an increase of the structural-magnetic transition at $T_{s,N}$ with increasing amount of Sr, with the transition staying first order. The evolution of $T_{s,N}$ and the pressure-induced non-magnetic collapsed tetragonal phase at $T_{cT}$ with pressure has been studied in detail for crystals with $x$ = 0.105 and 0.177. For $x$ = 0.105 we find a suppression of the structural magnetic transition with increasing pressure which becomes truncated at finite temperature by the high-pressure cT phase. In contrast, for $x$ = 0.177 the o/afm can be separated from the cT phase on the pressure axis. The results for $T_{s,N}$($x$) for the various concentration levels could be collapsed on the $T_{s,N}$ line for the $x$ = 0 system by employing a pressure shift $\Delta P$($x$). The fact that $\Delta P$($x$) grows linearly with $x$ demonstrates that as for the structural-magnetic transition, Sr-substitution acts as a chemical pressure. However, no such collapse is revealed for the collapsed tetragonal phase which becomes progressively shifted to higher pressures with increasing $x$. For $x$ = 0.177 our results suggest that the first-order structural-magnetic transition can be fully suppressed and that there is a finite pressure window before the cT phase is stabilized. In this pressure range evidence for superconductivity is found. Our study indicates that Sr-substitution in combination with hydrostatic pressure provides another route for stabilizing superconductivity in CaFe$_2$As$_2$. Key element in both approaches is the separation of the o/afm and the cT phase which is possible due to their different response to hydrostatic pressure.  

\begin{acknowledgments}
Work at Ames Laboratory (S. R, S. L. B., P. C. C.) was supported by the U.S. Department of Energy, Office of Basic Energy Science, Division of Materials Sciences and Engineering.  Ames Laboratory is operated for the U.S. Department of Energy by Iowa State University under Contract No. DE-AC02-07CH11358.  PCC also acknowledged support from the Alexander von Humboldt Foundation.
\end{acknowledgments}

\bibliographystyle{apsrev}
\bibliography{literatur}

\end{document}